# Wave -> Diffusion Transition in Microtubules

Mirosław Kozłowski* Janina Marciak-Kozłowska

Institute of Electron Technology, Al. Lotników 32/46, 02-668 Warszawa, Poland

* Corresponding author ,e-mail:MiroslawKozlowski@aster.pl

Abstract

In this paper the heat transport in microtubules ( MT) is investigated. When the dimension of the structure is of the order of the de Broglie wave length the transport phenomena must be analyzed within quantum mechanics. In this paper we developed the Dirac type thermal equation for MT .The solution of the equation-the temperature fields for electrons can be wave type or diffusion type depending on the dynamics of the scattering.

# 1. Introduction

The interaction of the laser pulses with MT is a very interesting and new field of investigation. In the biology the MT form the skeleton of the living cells  The exceptional properties of the mic include  the ballistic transport of the nerve impulses  Microtubules are long little tubes, a few nanometer in diameter. In the case of the MT lying within neurons,  we can find them from one end of the axons and the dendrities Their  action has a lot to do with the transport of the neurotransmitter chemicals along axons, and the growth of dendrites [1]

The breakthrough progress has been made recently in generation and detection of ultrashort, attosecond laser pulses with high harmonic generation technique [2,3]]. This is the beginning of the attophysics age in which many-electron dynamics will be investigated in real time.

In the existing new laser projects [4] the generation of 100 GW-level attosecond X-ray pulses is investigated. With ultra-short (attosecond) high energetic laser pulses the relativistic multi-electrons states can be generated.

For relativistic one electron state the Dirac equation is the master equation. In this paper we develop and solve the Dirac type thermal equation for multi-electron states generated in laser interaction with matter. In this paper the Dirac one dimensional thermal equation will be applied to study of the generation of the positron-electron pairs. It will be shown that the cross section is equal to the Thomson cross-section for the electron-electron scattering.

## 2. Derivation of the 1+1 dimensional Dirac equation for thermal processes in microtubules

As pointed in paper [5] spin-flip occurs only when there is more than one dimension in space. Repeating the discussion of deriving the Dirac equation  for the case of one spatial dimension, one easily finds that the Dirac matrices **α**  and  **β**  are reduced to 2x2 matrices that can be represented by the Pauli matrices ]. This fact simply implies that if there is only one spatial dimension, there is no spin. It should be instructive to show explicitly how to derive the 1+1 dimensional Dirac equation.

As discussed in textbooks [5,6] a wave equation that satisfies relativistic covariance in space-time as well as the probabilistic interpretation should have the form:

$$i\hbar \frac{\partial}{\partial t}\Psi(x,t) = \left[ c\boldsymbol{\alpha}\left(-i\hbar\frac{\partial}{\partial x}\right) + \boldsymbol{\beta} m_0 c^2 \right]\Psi(x,t). \tag{1}$$

To obtain the relativistic energy-momentum relation $E^2 = (pc)^2 + m_0^2 c^4$ we postulate that (1) coincides with the Klein-Gordon equation

$$\left[\frac{\partial^2}{\partial (ct)^2} - \frac{\partial^2}{\partial x^2} + \left(\frac{m_0 c}{\hbar}\right)^2\right] \Psi(x,t) = 0. \qquad (2)$$

By comparing (1) and (2) it is easily seen that $\alpha$ and $\beta$ must satisfy

$$\alpha^2 - \beta^2 = 1, \qquad \alpha\beta + \beta\alpha = 0. \qquad (3)$$

Any two of the Pauli matrices can satisfy these relations. Therefore, we may choose $\alpha = \sigma_x$ and $\beta = \sigma_z$ and we obtain:

$$i\hbar \frac{\partial}{\partial t} \Psi(x,t) = \left[c\sigma_x \left(-i\hbar \frac{\partial}{\partial x}\right) + \sigma_z m_0 c^2\right] \Psi(x,t), \qquad (4)$$

where $\Psi(x,t)$ is a 2-component spinor.

The Eq. (4) is the Weyl representation of the Dirac equation. We perform a phase transformation on $\Psi(x,t)$ letting $u(x,t) = \exp\left(\frac{imc^2 t}{\hbar}\right) \Psi(x,t)$. Call $u$'s upper (respectively, lower) component $u_+(x,t)$, $u_-(x,t)$; it follows from (4) that $u_\pm$ satisfies

$$\frac{\partial u_\pm(x,t)}{\partial t} = \pm c \frac{\partial u_\pm}{\partial x} + \frac{im_0 c^2}{\hbar}(u_\pm - u_\mp). \qquad (5)$$

Following the physical interpretation of the equation (5) it describes the relativistic particle (mass $m_0$) propagates at the speed of light $c$ and with a certain *chirality* (like a two component neutrino) except that at random times it flips both direction of propagation (by $180^0$) and chirality.

In monograph[ 7] we considered a particle moving on the line with fixed speed $w$ and supposed that from time to time it suffers a complete reversal of direction, $u(x,t) \Leftrightarrow v(x,t)$, where $u(x,t)$ denotes the expected density of particles at $x$ and at time $t$ moving to the right, and $v(x,t) \equiv$ expected density of particles at $x$ and at time $t$ moving to the left. In the following we perform the change of the abbreviation

$$\begin{aligned} u(x,t) &\to u_+, \\ v(x,t) &\to u_-. \end{aligned} \qquad (6)$$

Following the results of the paper [6] we obtain for the $u_\pm(x,t)$ the following equations

$$\frac{\partial u_+}{\partial t} = -w\frac{\partial u_+}{\partial x} - \frac{w}{\lambda}\left((1-k)u_+ - ku_-\right),$$
$$\frac{\partial u_-}{\partial t} = w\frac{\partial u_-}{\partial x} + \frac{w}{\lambda}\left(ku_+ + (k-1)u_-\right). \quad (7)$$

In equation (7) $k(x)$ denotes the number of the particles which are moving in left (right) direction after the scattering at $x$. The mean free path for scattering is equal $\lambda$, $\lambda = w\tau$, where $\tau$ is the relaxation time for scattering.

Comparing formulae (5) and (7) we conclude that the shapes of both equations are the same. In the subsequent we will call the set of the equations (7) *the Dirac equation* for the particles with velocity $w$, mean free path $\lambda$.

For thermal processes we define $T_{+,-} \equiv$ the temperature of the particles with chiralities + and – respectively and with analogy to equation (7) we obtain:

$$\frac{\partial T_+}{\partial t} = -w\frac{\partial T_+}{\partial x} - \frac{w}{\lambda}\left((1-k)T_+ - kT_-\right),$$
$$\frac{\partial T_-}{\partial t} = w\frac{\partial T_-}{\partial x} + \frac{w}{\lambda}\left(kT_+ + (k-1)T_-\right), \quad (8)$$

where $\frac{w}{\lambda} = \frac{1}{\tau}$.

In one dimensional case we introduce one dimensional cross section for scattering

$$\sigma(x,t) = \frac{1}{\lambda(x,t)}. \quad (9)$$

### 3. The solution of the Dirac equation for stationary temperatures in microtubules

In the stationary state thermal transport phenomena $\frac{\partial T_{+,-}}{\partial t} = 0$ and Eq. (8) can be written as

$$\frac{dT_+}{dx} = -\sigma\left((1-k)T_+ + kT_-\right),$$
$$\frac{dT_-}{dx} = \sigma(k-1)T_- + \sigma k T_+. \quad (10)$$

After the differentiation of the equation (9) we obtain for $T_+(x)$

$$\frac{d^2 T_+}{dx^2} - \frac{1}{\sigma k}\frac{d}{dx}(\sigma k)\frac{dT_+}{dx} + T_+\left[\sigma^2(2k-1) + \frac{d\sigma}{dx}(1-k) + \frac{\sigma(k-1)}{\sigma k}\frac{d(\sigma k)}{dx}\right] = 0.$$

Equation (10) can be written in a compact form

$$\frac{d^2 T_+}{dx^2} + f(x)\frac{dT_+}{dx} + g(x)T_+ = 0,$$

where

$$f(x) = -\frac{1}{\sigma}\left(\frac{\sigma}{k}\frac{dk}{dx} + \frac{d\sigma}{dx}\right),$$
$$g(x) = \sigma^2(x)(2k-1) - \frac{\sigma}{k}\frac{dk}{dx}. \tag{11}$$

In the case for constant $\frac{dk}{dx} = 0$ we obtain

$$f(x) = -\frac{1}{\sigma}\frac{d\sigma}{dx},$$
$$g(x) = \sigma^2(x)(2k-1). \tag{12}$$

With functions $f(x)$, $g(x)$ described by formula (12) the general solution of Eq. (12) has the form:

$$T_+(x) = C_1 e^{(1-2k)^{\frac{1}{2}}\int \sigma(x)dx} + C_2 e^{-(1-2k)^{\frac{1}{2}}\int \sigma(x)dx} \tag{13}$$

and

$$T_-(x) = \frac{\left[(1-k)+(1-2k)^{\frac{1}{2}}\right]}{k} \times$$
$$\left[C_1 e^{(1-2k)^{\frac{1}{2}}\int \sigma(x)dx} + \frac{(1-k)-(1-2k)^{\frac{1}{2}}}{(1-k)+(1-2k)^{\frac{1}{2}}} C_2 e^{-(1-2k)^{\frac{1}{2}}\int \sigma(x)dx}\right]. \tag{14}$$

The formulae (13) and (14) describe three different mode for heat transport. For $k = \frac{1}{2}$ we obtain $T_+(x) = T_-(x)$ while for $k > \frac{1}{2}$, i.e. for heat carrier generation $T_+(x)$ and $T_-(x)$ are the thermal waves. For $k < \frac{1}{2}$ i.e. for strong absorption $T_+(x)$ and $T_-(x)$ represents the diffusion motion. The mechanism of the scattering is for the moment unknown. The farreaching hypothesis is that the electrons interact with the point field (ZPF) and as the result the additional carriers are generated

In the subsequent we will consider the solution of Eq. (9) for Cauchy conditions:
$$T_+(0) = T_0, \quad T_-(a) = 0. \tag{15}$$

Boundary conditions (15) describes the generation of heat carriers by illuminating the left end of one dimensional slab (with length $a$) by laser pulse.

From formulae (13) and (14) we obtain:

$$T_+(x) = \frac{2T_0 e^{[f(0)-f(a)]}}{1+\beta e^{2[f(0)-f(a)]}} \times \frac{(1-2k)^{\frac{1}{2}}\cosh[f(x)-f(a)] + (k-1)\sinh[f(x)-f(a)]}{(1-2k)^{\frac{1}{2}} - (k-1)}, \tag{16}$$

$$T_-(x) = \frac{2T_0 e^{2[f(0)-f(a)]} \left[(k-1)+(1-2k)^{\frac{1}{2}}\right] \sinh[f(x)-f(a)]}{(1+\beta e^{-2[f(a)-f(0)]})k}. \quad (17)$$

In formulae (16) and (17)

$$\beta = \frac{(1-2k)^{\frac{1}{2}}+(k-1)}{(1-2k)^{\frac{1}{2}}-(k-1)} \quad (18)$$

and

$$f(x) = (1-2k)^{\frac{1}{2}} \int \sigma(x)dx,$$
$$f(0) = (1-2k)^{\frac{1}{2}} \left[\int \sigma(x)dx\right]_0, \quad (19)$$
$$f(a) = (1-2k)^{\frac{1}{2}} \left[\int \sigma(x)dx\right]_a.$$

With formulae (16) and (17) for $T_+(x)$ and $T_-(x)$ we define the asymmetry $A(x)$ of the temperature $T(x)$

$$A(x) = \frac{T_+(x)-T_-(x)}{T_+(x)+T_-(x)}, \quad (20)$$

$$A(x) = \frac{\dfrac{(1-2k)^{\frac{1}{2}}}{(1-2k)^{\frac{1}{2}}-(k-1)}\cosh[f(x)-f(a)]-\dfrac{1-2k}{(1-2k)^{\frac{1}{2}}-(k-1)}\sinh[f(x)-f(a)]}{\dfrac{(1-2k)^{\frac{1}{2}}}{(1-2k)^{\frac{1}{2}}-(k-1)}\cosh[f(x)-f(a)]-\dfrac{1}{(1-2k)^{\frac{1}{2}}-(k-1)}\sinh[f(x)-f(a)]} \quad (21)$$

From formula (21) we conclude that for elastic scattering, i.e. when $k = \dfrac{1}{2}$, $A(x) = 0$, and for $k \neq \dfrac{1}{2}$, $A(x) \neq 0$.

In the monograph [7] we introduced the relaxation time $\tau$ for quantum heat transport

$$\tau = \frac{\hbar}{mv^2}. \quad (22)$$

In formula (22) $m$ denotes the mass of heat carriers electrons and $v = \alpha c$, where $\alpha$ is the fine structure constant for electromagnetic interactions. As was shown in monograph [7], $\tau$ is also the lifetime for positron-electron pairs in vacuum.

When the duration time of the laser pulse is shorter than $\tau$, then to describe the transport phenomena we must use the hyperbolic transport equation. Recently the structure of water was investigated with the attosecond $(10^{-18} \text{s})$ resolution [8]. Considering that $\tau \approx 10^{-17}$ s we argue that to study performed in [8] open the new field for investigation of laser pulse with matter. In order to apply the equations (9) to attosecond laser induced phenomena we must know the cross section $\sigma(x)$. Considering formulae (9) and (22) we obtain

$$\sigma(x) = \frac{mv}{\hbar} = \frac{me^2}{\hbar^2} \qquad (23)$$

and it occurs $\sigma(x)$ is the Thomson cross section for electron-electron scattering.

With formula (23) the solution of Cauchy problem has the form:

$$T_+(x) = \frac{2T_0 e^{-(1-2k)^{\frac{1}{2}}\frac{me^2}{\hbar^2}a}}{\left[1 + \beta e^{-2(1-2k)^{\frac{1}{2}}\frac{me^2}{\hbar^2}a}\right]} \times$$

$$\frac{(1-2k)^{\frac{1}{2}}\cosh\left[(1-2k)^{\frac{1}{2}}\frac{me^2}{\hbar^2}(x-a)\right] + (k-1)\sinh\left[(1-2k)^{\frac{1}{2}}\frac{me^2}{\hbar^2}(x-a)\right]}{(1-2k)^{\frac{1}{2}} - (k-1)},$$

$$T_-(x) = \frac{2T_0 e^{-\frac{(1-2k)^{\frac{1}{2}}me^2 a}{\hbar^2}}\left[(k-1) - (1-2k)^{\frac{1}{2}}\right]}{\left(1 + \beta e^{-2(1-2k)^{\frac{1}{2}}\frac{me^2}{\hbar^2}a}\right)k} \times$$

$$\sinh\left[(1-2k)^{\frac{1}{2}}\frac{me^2}{\hbar^2}(x-a)\right].$$

(24)

In this paper we investigate the interaction of the ultra short laser pulses with MT`s of the length of the 25 nm.

In Figs. 1-3 the results of calculations are presented. All the calculations are performed for microtubule with length a= 25 nm. In Fig.1 the result for A[x,k] is obtained when mean free path, l = 1nm. In Fig. 2 the mean free path is assumed l =10 nm. The shape of A[x,k] for small value k, 0<k<0.5 is presented in Fig.3. As can be concluded for k ~ 0.5 the transition from wavy motion to the diffusion is observed.

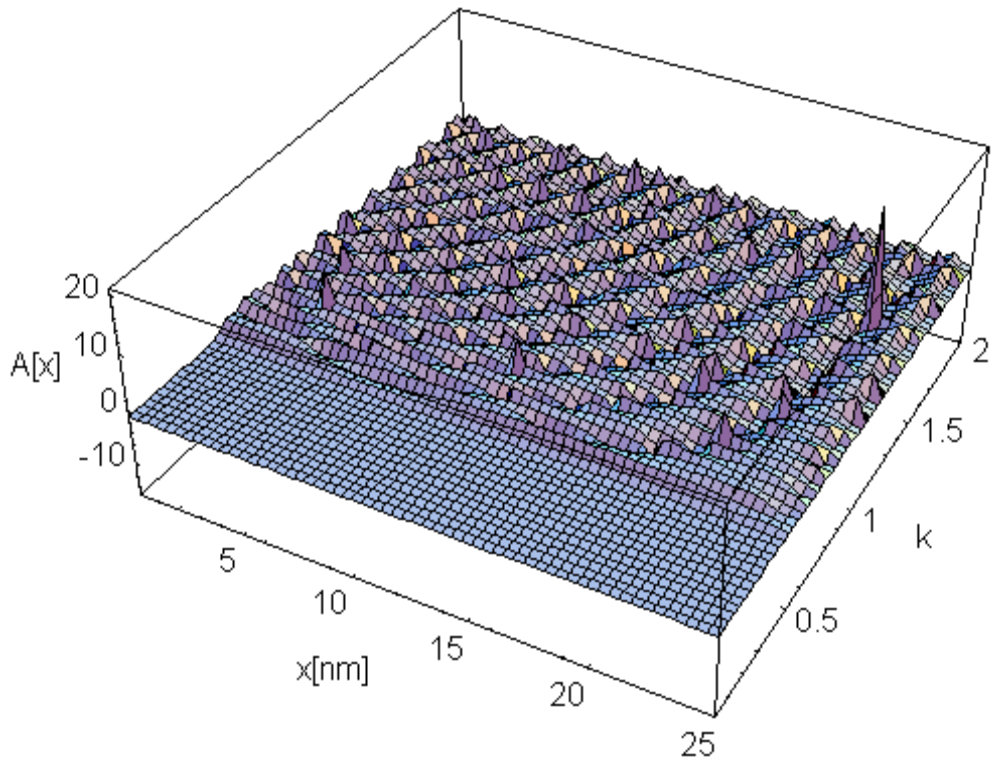

Fig.1, a=25 nm, l=1 nm

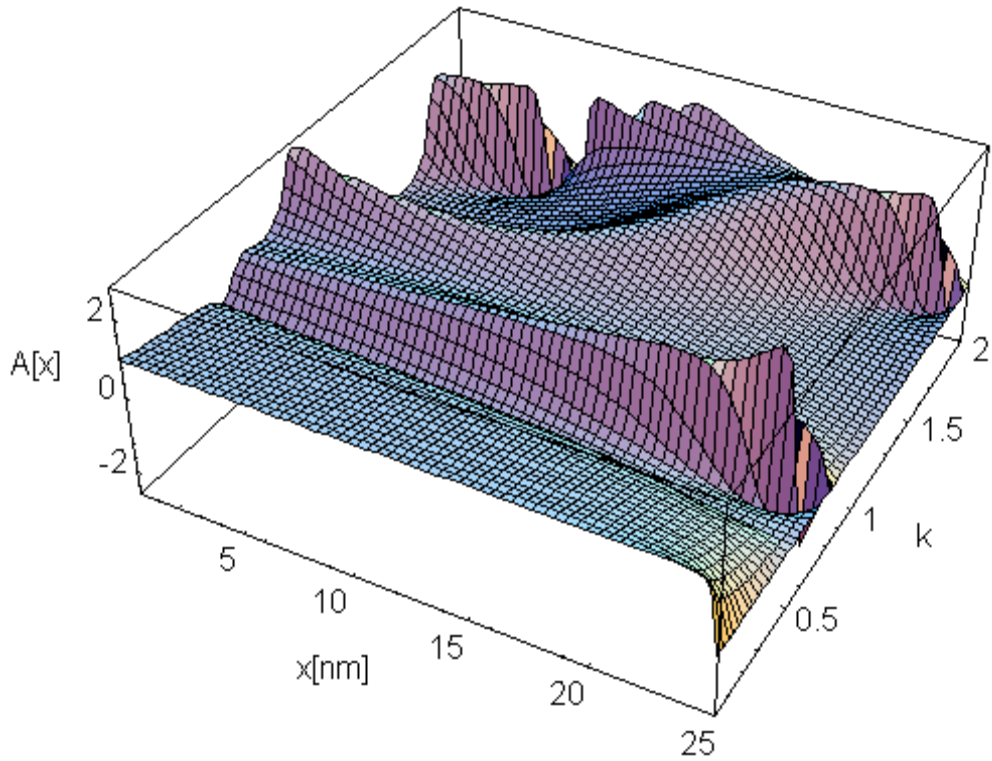

Fig.2, a=25 nm, l=10 nm

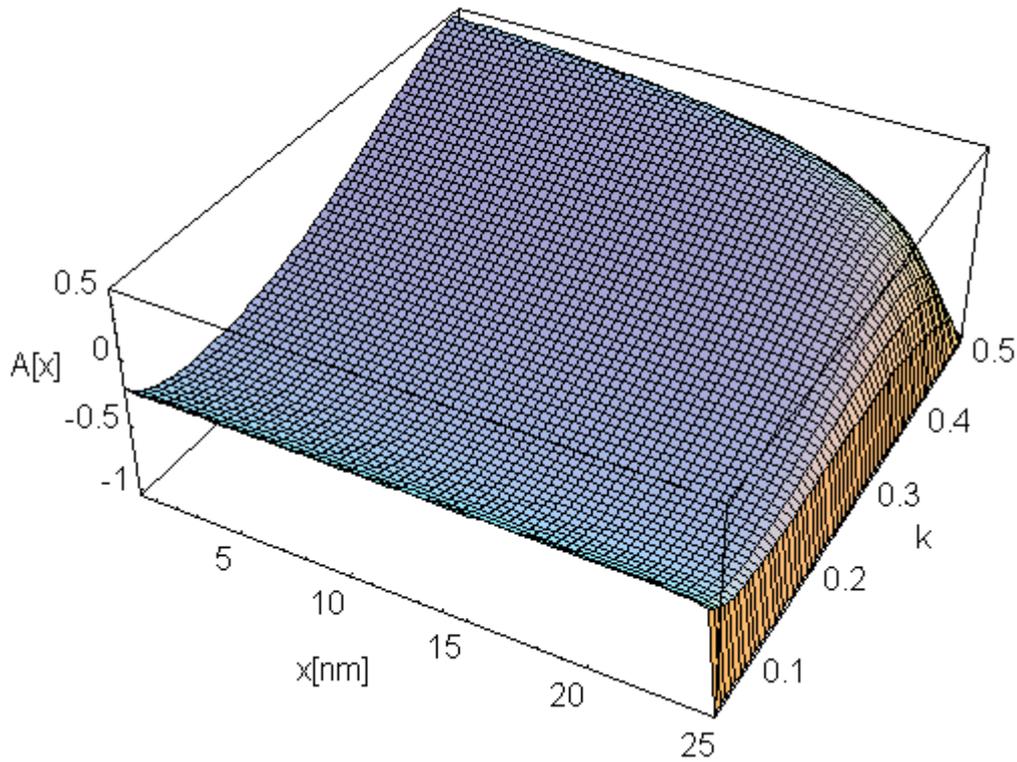

**4 Conclusions**

In this paper the one dimensional Dirac type thermal equation for MT was developed and solved. It was shown that depending on the dynamics of the heat carriers scattering the wave or diffusion temperature fields can be generated in MT.